\documentclass[lettersize,journal]{IEEEtran}
\usepackage{amsmath,amsfonts}
\usepackage{algorithmic}
\usepackage{algorithm}
\usepackage{array}
\usepackage[caption=false,font=normalsize,labelfont=sf,textfont=sf]{subfig}
\usepackage{textcomp}
\usepackage{stfloats}
\usepackage{url}
\usepackage{verbatim}
\usepackage{graphicx}
\usepackage{cite}
\usepackage{hyperref}
\usepackage{multirow}
\usepackage{pifont} 
\usepackage{amsmath}
\usepackage{float}

\begin{document}

\title{PreAdaptFWI: Pretrained-Based Adaptive Residual Learning for Full-Waveform Inversion Without Dataset Dependency}

\author{Xintong Dong, Zhengyi Yuan, Jun Lin, Shiqi Dong, Xunqian Tong, Yue Li,
\thanks{This work was supported in part by the National
Natural Science Foundation of China under Grant 42230805, 42204114, 42374222, and  Doctoral Research Foundation of Northeast Electric Power University under Grant BSJXM-2023203. (Corresponding author: Shiqi Dong , Xunqian Tong)}
\thanks{Xintong Dong, Zhengyi Yuan, Jun Lin and Xunqian Tong are with the College of Instrumentation and Electrical Engineering, Jilin University, Changchun 130026, China. (e-mail: dxt@jlu.edu.cn, yuanzhengyi0224@163.com, lin\_jun@jlu.edu.cn, txq@jlu.edu.cn)}
\thanks{Shiqi Dong are with the Key Laboratory of Modern Power System Simulation and Control and Renewable Energy Technology, Northeast Electric Power University, Jilin 132012, China. (e-mail: dsq1994@126.com)}
\thanks{Yue Li are with the College of Communication Engineering, Jilin University, Changchun, 130012, China. (e-mail: liyue@jlu.edu.cn)}
}

\markboth{Journal of \LaTeX\ Class Files,~Vol.~14, No.~8, August~2021}%
{Shell \MakeLowercase{\textit{et al.}}: A Sample Article Using IEEEtran.cls for IEEE Journals}


\maketitle

\begin{abstract}
Full-waveform inversion (FWI) is a method that utilizes seismic data to invert the physical parameters of subsurface media by minimizing the difference between simulated and observed waveforms. Due to its ill-posed nature, FWI is susceptible to getting trapped in local minima. Consequently, various research efforts have attempted to combine neural networks with FWI to stabilize the inversion process. This study presents a simple yet effective training framework that is independent of dataset reliance and requires only moderate pre-training on a simple initial model to stabilize network outputs. During the transfer learning phase, the conventional FWI gradients will simultaneously update both the neural network and the proposed adaptive residual learning module, which learns the residual mapping of large-scale distribution features in the network's output, rather than directly fitting the target mapping. Through this synergistic training paradigm, the proposed algorithm effectively infers the physically-informed prior knowledge into a global representation of stratigraphic distribution, as well as capturing subtle variations in inter-layer velocities within local details, thereby escaping local optima. Evaluating the method on two benchmark models under various conditions, including absent low-frequency data, noise interference, and differing initial models, along with corresponding ablation experiments, consistently demonstrates the superiority of the proposed approach.

\end{abstract}


\section{Introduction}
\IEEEPARstart{S} eismic inversion plays a crucial role in the study of Earth's internal structure and the development of natural resources. Therefore, accurate velocity model building is a key task for seismic imaging and interpretation.  Traditional inversion methods, such as migration velocity analysis and tomography, rely solely on the kinematic information from seismic data, which limits their ability to precisely capture the fine-scale structures of subsurface media. The introduction of Full-Waveform Inversion\cite{1,2,3,4} provides a novel approach, where simulated data predicted by solving the acoustic or elastic wave equation is iteratively matched with seismic records. This process updates the model until an optimal solution is reached. However, due to the ill-posed nature of the inversion process and the fact that this algorithm typically relies on the adjoint state method\cite{5} to compute the gradient of the objective function with respect to the unknown variables, the phase difference between the simulated and observed waveforms during waveform fitting can easily exceed half a period. This phenomenon, known as cycle skipping, leads to incorrect fitting of waveforms with different periods, ultimately causing the algorithm to get trapped in local minima\cite{6}.

Obtaining a good initial model and rich low-frequency data is considered the optimal solution to address the issue of local minima. However, due to practical challenges in exploration, the acquisition of such data often does not align with ideal conditions. As a result, researchers have begun exploring various approaches to tackle the cycle skipping problem that arises in the absence of a good initial model and abundant low-frequency data. Some studies have attempted to use a hierarchical multi-scale strategy\cite{7,8,9,10,11,12}, where model parameters are iteratively inverted by decomposing the scales from low frequency to high frequency. 

However, in real-world field data acquisition, low-frequency signals are often insufficient or contaminated by noise, which hampers the provision of low-wavenumber components for the model parameters, posing a significant challenge in the initial stages of Full-Waveform Inversion. To address this issue, Many researchers have explored transforming seismic data into the Laplace domain or the Laplace-Fourier domain for analysis and processing\cite{13,14,15,16,17,18,19,20}. Other studies\cite{21,22,23,24} have attempted to recover low-frequency information by extracting the envelope of seismic data, thereby enhancing subsurface structure identification. Meanwhile, several studies have focused on using regularization techniques\cite{25,26,27,28,29,30,31} to reduce the ill-posedness during the inversion process. 

\begin{figure}[t]
\label{f1}
	\centering
	\includegraphics[width=\linewidth]{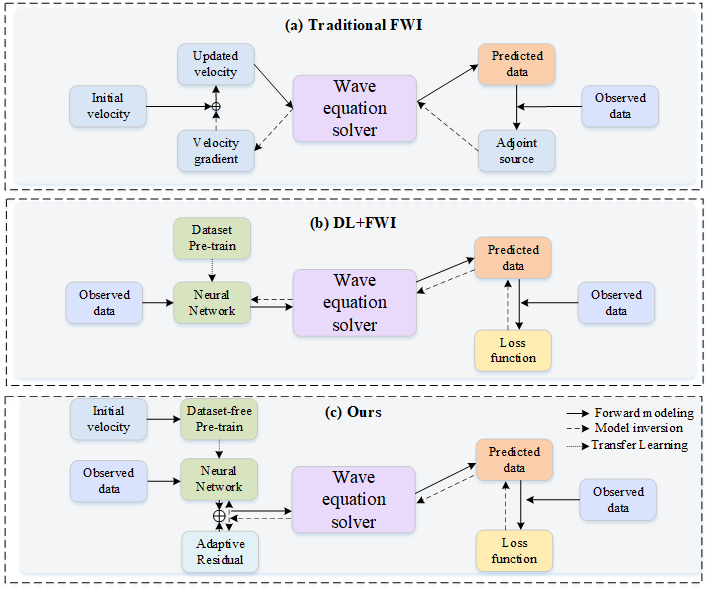}
	\caption{Overview of previous FWI methods that are closely related to our approach.}
\end{figure}

The introduction of the above regularization methods has improved the stability and accuracy of the inversion results. However, when sharp boundaries or discontinuous structures exist in the model, the regularization terms may smooth the velocity model, potentially compromising the inversion accuracy. Moreover, it is noteworthy that the inversion efficiency can significantly decrease due to the repeated adjustments of the regularization parameters during the iterative process, coupled with the additional computational burden. This makes it challenging to meet the demands for speed and efficiency in large-scale data processing or time-sensitive real-world applications. Furthermore, in cases where the model structure is complex and contains many discontinuous features, regularization methods may cause the inversion to converge to a local optimum, preventing the accurate capture of the model's true characteristics. This further limits their applicability and effectiveness in inversion problems involving complex geological structures. Recently, deep learning-based methods have been widely explored in geophysical problems \cite{32,33,34,35,36,37,38,39,40,41,42,43,44,45,46,47,48,49,50}, Yang and Ma\cite{49} proposed a supervised deep fully convolutional neural network method based on U-Net, which directly reconstructs velocity models from prestack data. This data-driven approach automatically extracts multi-level features without manual intervention, significantly reducing computation time, and achieved excellent results on a salt dome velocity model. 

\begin{figure*}[t]
\label{f2}
	\centering
	\includegraphics[width=\linewidth]{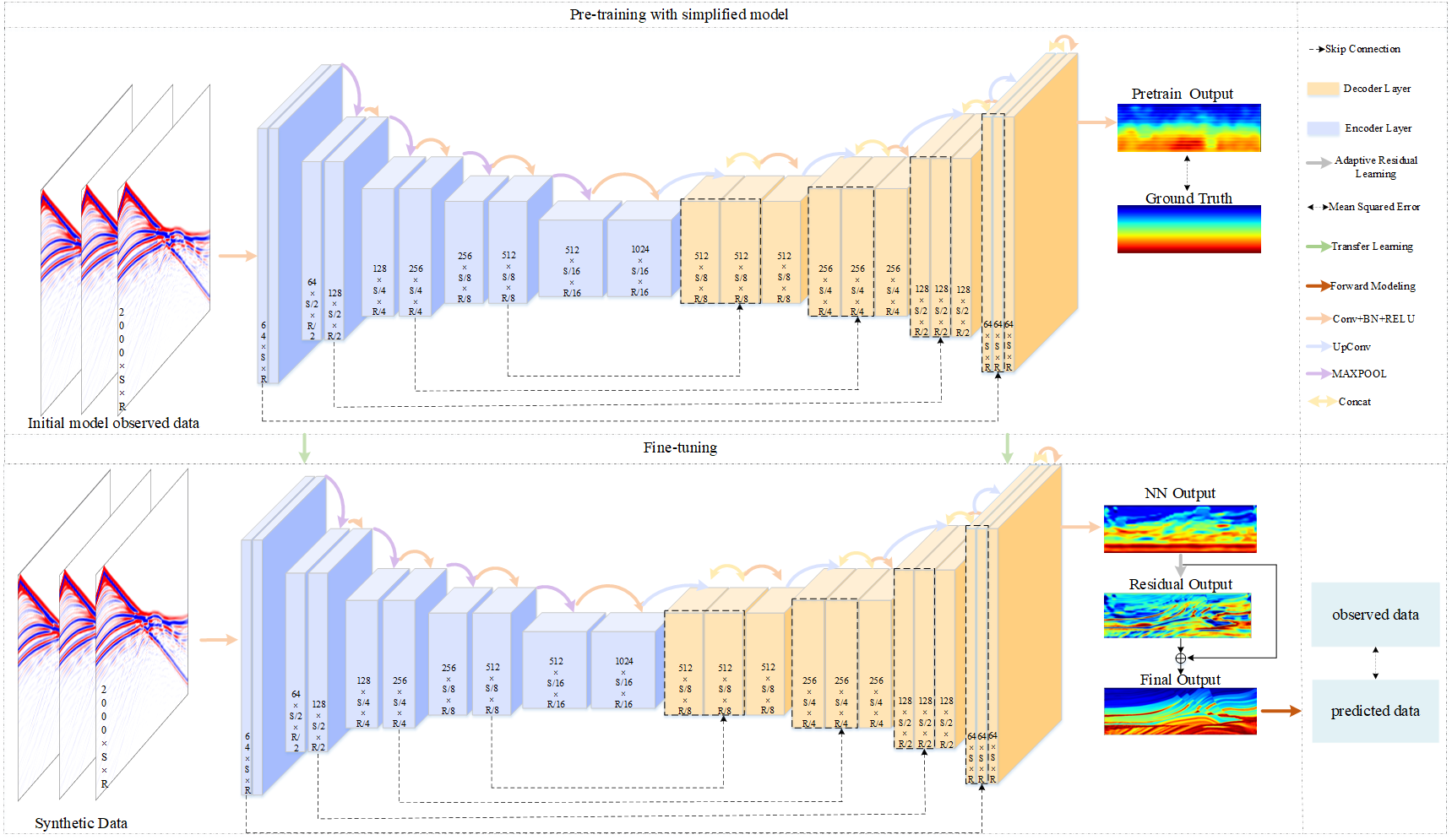}
	\caption{The schematic workflow of the proposed PreAdaptFWI. After pre-training with the observed data corresponding to a simple initial model, transfer learning is employed to achieve end-to-end prediction using real observational data.}
\end{figure*}

However, the training performance of convolutional neural networks depends on the data-driven supervised learning approach. Due to the difficulty and variability in obtaining seismic data, neural networks often reach suboptimal solutions because of the lack of sufficient real labeled data for training. Therefore, a common approach is to use synthetic data for training the neural networks\cite{51}. Muller et al. \cite{52} proposed a method combining supervised learning and physics-guided neural networks for full-waveform inversion, where a pre-trained CNN is used to initiate the inversion process through transfer learning, significantly reducing uncertainties and accelerating convergence with minimal prior information. Guo et al. addressed the limitations of existing datasets by creating enhanced datasets with varying source locations and frequencies. Their proposed Inversion-DeepONet\cite{53} architecture utilizes convolutional neural networks to extract features from seismic data, incorporating source parameters such as locations and frequencies. In this context, researchers are required to manually design datasets to facilitate network fitting. However, this approach often leads to a training process that heavily relies on manually selected features and data, limiting the network's adaptability and potentially increasing the cost and complexity of data preparation\cite{54,55,56}.
 
Another area of research\cite{57,58,59,60,61,62,63,64,65} attempts to integrate physical laws with neural networks. The introduction of prior knowledge imposes strong constraints on the network, addressing the need for large amounts of training data in supervised learning while improving the stability of the network during the training process. The data-driven approach in deep learning relies on the assumption that high performance can be achieved by capturing the extensive variability in the input space through large and diverse datasets. However, neural networks trained with synthetic data often face challenges in generalization and domain adaptation when applied to real-world data, such as issues with waveform matching to observed data and dependency on velocity features during the training process. Physics-informed regularization methods, which constrain and optimize the model through wave equations, guide the algorithm away from unreasonable solutions and mitigate the occurrence of local minima. However, these methods require the incorporation of prior knowledge, which limits the flexibility of the model. 

Inspired by the challenges and limitations faced in the aforementioned works, we propose a novel pretrained-based adaptive residual learning neural network for full-waveform inversion. Specifically, the contributions of this paper are as follows: (1) This paper designs a simple yet effective training framework. The key distinction from other methods is that this framework only involves appropriate pretraining on a simple initial model, meaning that it stabilizes the neural network training process without the need for manually constructed datasets. This ensures that the proposed method not only possesses strong nonlinear fitting capabilities but also fundamentally eliminates the issue of domain mismatch. Domain mismatch occurs when a neural network trained on synthetic datasets learns a feature space that differs from that of real-world data. Specifically, synthetic data often fails to fully capture the true variability and complexity of real-world scenarios, leading to a discrepancy between the features learned from synthetic data and those present in actual observed data. This gap can significantly limit the model's ability to generalize to real-world applications. By addressing this issue, the proposed method ensures a closer alignment between the learned features and real data, enhancing the model's performance and reliability. (2) We simultaneously update the neural network and the proposed adaptive residual learning module using FWI gradients that adhere to physical constraints. The introduction of this module facilitates the separation of global and fine-grained features. Through bidirectional optimization, the physical constraints are integrated, allowing the neural network to focus on learning large-scale velocity distributions, while the adaptive residual module captures subtle local velocity variations. This approach enables synergistic complementarity between different features, significantly enhancing the network’s ability to interpret complex geological structures, and ultimately converging to the optimal solution.

\section{Method}
\subsection{Traditional FWI}
Full-waveform inversion is a widely applied inversion technique primarily employed to infer subsurface velocity models. It simulates the propagation of the wavefield through an initial model using the wave equation $F(w,p)$, where $w$ represents the wavefield and $p$ denotes the model parameters of the medium. The wavefield is obtained by solving the forward wave equation, providing information about wave propagation and describing the temporal evolution of seismic waves in the subsurface. The computational process is formulated as follows:
\begin{equation}
    w(x,z,t)=F(p)
\end{equation}
$F$ represents the forward modeling operator of the wave equation, while the variables, $x$, $z$, and $t$ correspond to the lateral distance, subsurface depth, and temporal evolution of wave propagation, respectively. The next step is to extract the simulated data $d{}_{sim}$ from the computed wavefield:
\begin{equation}
    d{}_{sim}=S{}_{w}(x,z,t)
\end{equation}
$S$ is the observation operator, representing the extraction of the wavefield $w$ at the receiver locations.

FWI estimates velocity model parameters $p$ that best align with the observed seismic wavefield $d{}_{obs}$. This alignment is quantified using an appropriate objective function
$O(d{}_{obs},d{}_{sim})$ that measures the fit between the observed and simulated data. The model parameters are determined as the solution to this optimization problem.

Therefore, how to minimize the objective function subject to the constraints of equations (1) and (2). The model parameters $p$ are adjusted through linear iterations. At each iteration, the model $p$ is updated based on the gradient $\frac{\partial O}{\partial P}$ and step size $\alpha$. The iteration terminates when the objective function falls below the specified threshold or when the maximum allowable number of iterations is reached.

\begin{equation}
\label{e3}
p{}_{i+1}=p{}_{i}-\alpha \frac{\partial O}{\partial{p}_{{}_{i}}}
\end{equation}

Since FWI is an ill-posed problem, the accuracy of the solution is highly dependent on the initial model. When the difference between the initial and true models causes the phase difference between the observed data and the simulated data to exceed half a wavelength, cycle skipping can arise. Therefore, reducing the reliance on the initial model and minimizing the likelihood of cycle skipping have become key research focuses in the field.

\subsection{Deep learning driven FWI}

The method proposed in this paper is primarily inspired by recent works that combine deep learning with traditional FWI techniques, with the process outlined in Fig 1. Seismic data are mapped to the model via a neural network, which is then employed as input for traditional FWI. The corresponding equation is as follows:

\begin{equation}
\label{e4}
p=Conv(d{}_{sim},\theta )
\end{equation}
\begin{equation}
\label{e5}
\theta {}_{i+1}=\theta {}_{i}-\alpha \frac{\partial p}{\partial{\theta }_{{}_{i}}}\cdotp \frac{\partial O}{\partial{p}_{{}_{i}}}
\end{equation}
where $Conv$ denotes the convolutional neural network that transforms the simulated data from the data domain to the spatial domain, and $\theta$ represents the set of all learnable weight parameters used in this process.

Providing seismic data as input to convolutional neural networks can effectively integrate background knowledge and serve as a robust constraint mechanism. Without supervised training to optimize the weights in accordance with the tasks represented by the model, CNN weights are typically initialized randomly. During the initial iteration phases, models generated using neural network methodologies tend to exhibit inconsistent shapes and substantial deviations from the true model.

In Deep pre-trained FWI\cite{52}, the incorporation of pre-trained transfer learning methods effectively mitigates the uncertainty associated with random weight initialization. Subsequently, during the early stages of inversion, the trained network weights are employed as the initial values for the network. These weights are then iteratively refined under the guidance of the FWI gradient. However, it is noteworthy that the proposed approach relies on manually curated synthetic datasets, and although effective in specific experimental contexts, these datasets exhibit substantial discrepancies when compared to real seismic data. This discrepancy limits the practical applicability of the method in real-world scenarios. Furthermore, the manual creation of datasets is both time-consuming and resource-intensive, which further restricts the scalability and broader implementation of the approach.

\subsection{PreAdaptFWI}
This paper proposes a dataset-independent pretraining framework for FWI, which utilizes a pretrained CNN to represent the velocity model, replacing the need for randomly initialized weights. Unlike previous approaches, this method does not require synthetic datasets for pretraining. Instead, the linear initial model, denoted as $p{}_{simple}$, is used to generate seismic data $d{}_{simple}$, which serve as the labels and inputs during the pretraining process. This process is represented by the function $p{}_{pred}=CNN(d{}_{simple},\theta )$, where the generated $p{}_{pred}$ is compared with the label data to iteratively adjust the neural network weights. The pretrained neural network weight is then employed as the initial weights for inversion, a process commonly referred to as transfer learning. The proposed method demonstrates a significant improvement in inversion performance, even within an overly simplified parameter space.

In the selection of the neural network architecture, U-Net\cite{66} was chosen as the model framework. Originally developed for medical image segmentation tasks, U-Net's distinctive encoder-decoder structure is highly effective at capturing spatial hierarchical features in images. This capability is particularly crucial for the model updating process in seismic inversion. Consequently, U-Net has become a widely accepted and extensively utilized architecture in seismic science research\cite{49,52,62,65,67}. The Unet framework employed in this study is presented in detail in Fig 2. 

Seismic shot data serves as the input feature to the network, with each shot corresponding to one channel dimension of the input image. The receivers are arranged in a fixed configuration, with one receiver at each horizontal grid position. As a result, the horizontal size of the feature dimension aligns with the velocity model, while the vertical size corresponds to the number of recorded shots. This configuration leads to an imbalance in the feature dimension ratio, with a significantly larger amount of data in the vertical dimension compared to the horizontal dimension. 

The seismic source data used in this study has a recording duration of 6 seconds with a sampling rate of 0.003 seconds, resulting in 2,000 vertical sampling points. These data points are input as channel dimensions into the encoder of the U-Net for feature extraction. The encoder performs hierarchical downsampling of the input data through four convolutional modules, enabling the extraction of both global and local features. Each convolutional module is followed by a max-pooling layer, which reduces the horizontal and vertical dimensions by half. At the end of the encoder, a bottleneck layer composed of convolutional operations is employed to perform feature extraction and compression at deeper layers of the network, capturing high-level abstract features of the input data. The most significant stratigraphic structural information extracted is used as a reference for restoring spatial resolution during the decoder phase. The encoder-decoder section consists of four convolutional modules and upsampling layers. Through gradual upsampling and convolution operations, global features are combined with local features from the encoder output via skip connections. This approach ensures that both fine-grained details and global structural  information are preserved during the reconstruction process, ultimately producing a result that is both accurate and aligned with the size of the velocity model. The U-Net implementation in this study contains 32,224,793 trainable parameters, which accounts for only 10\% of the parameters in the original U-Net model, significantly reducing the parameter count.

\begin{figure}[t]
	\centering
	\includegraphics[width=\linewidth]{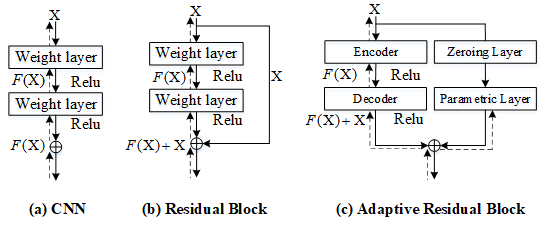}
	\caption{Comparison of the proposed adaptive residual learning module with conventional convolutional modules and residual learning modules.}
\end{figure}

In the process of constructing the velocity model using a convolutional neural network, a finite-difference method is employed to solve the acoustic wave equation, generating a small batch of seismic data. The adoption of a mini-batch strategy enhances training efficiency by allowing the model to optimize using smaller subsets of data in each iteration, thereby reducing computational resource requirements and accelerating convergence. The mini-batch size is set to a single seismic trace, which is randomly selected for each model update. 

The FWI gradient is computed using the adjoint-state method. However, rather than directly applying the gradient to update the velocity model, it is simultaneously propagated to the CNN and an adaptive residual module to adjust the weights of learnable parameters. Subsequently, the CNN, leveraging the optimized weights, generates an updated velocity model that is expected to be closer to the true model than its predecessor. This is a joint optimization process, where each time the CNN generates a new model, it is evaluated within the FWI framework using the adjoint-state method to compute the gradient. The gradient is not only used to update the CNN weights but is also passed to the adaptive residual module to adjust its learnable parameters, thereby refining the model more accurately. Throughout the inversion process, both the network weights and residual parameters are iteratively optimized based on the input seismic data and the physical principles of wave propagation, progressively approaching the true velocity model. The specific process of PreAdaptFWI is illustrated by the following formula:

\begin{equation}
X_{\text{enc}}^{(l)} = \text{MaxPool} \left( \sigma \left( W_{\text{enc}}^{(l)} * d_{\text{sim}} + b_{\text{enc}}^{(l)} \right) \right)
\end{equation}

\begin{equation}
X_{\text{dec}}^{(l)} = \sigma \left( W_{\text{dec}}^{(l)} * \text{Concat} \left( X_{\text{up}}^{(l)}, X_{\text{enc}}^{(L-l)} \right) + b_{\text{dec}}^{(l)} \right)
\end{equation}

\begin{equation}
    p=Conv{}_{1\times 1}(X_{\text{dec}}^{(l)})
\end{equation}

\begin{equation}
\begin{aligned}
p_{Adapt} &= \text{ARLM}(p) \\
          &= \text{Para}(\text{Zero}(p))
\end{aligned}
\end{equation}

\begin{equation}
    p{}_{Final}=p\oplus p_{Adapt}
\end{equation}

The simulated data $d_{sim}$ obtained from Equation 2 is used as the input to the U-Net network for feature extraction. At each layer of the encoding path, the feature map $X^{(l)}$ at layer $l$ undergoes feature extraction through a convolutional layer, followed by activation function processing. Subsequently, max pooling is applied to reduce the resolution. Here, $*$ represents the convolution operation, while $W^{(l)}$ and $b^{(l)}$ denote the convolution kernel weights and bias term, respectively, with $l \in \left\{0,1,2,3\right\}$. The function $\sigma(\cdot)$ represents the ReLU activation function, and MaxPool denotes the max pooling operation.

In the decoding path, each layer first performs upsampling, then concatenates with the skip connections from the encoding path, followed by feature extraction through convolutional layers. Here, $X_{\text{up}}^{(l)}$ represents the feature map after upsampling, $X_{\text{enc}}^{(l)}$ is the feature map at the corresponding layer of the encoding path, and $Concat(\cdot)$ denotes the concatenation operation along the channel dimension. Finally, U-Net employs $Conv_{1 \times 1}$ convolution to map the channel dimensions to the target dimension, generating the prediction results $p$, which are subsequently fed into the Adaptive Residual Learning Module (ARLM) to obtain the residual learning feature $p_{Adapt}$. The $Zero$ operation represents the zeroing layer, and the $Para$ operation is the parametric layer. It is noteworthy that the zeroing and parametric operations are only performed once during the training process. The final velocity model $p_{Final}$ is obtained by adding the outputs from the U-Net network and the adaptive residual learning module along the spatial dimension, where $\oplus$ denotes the addition operation.

PreAdaptFWI feeds the iteratively generated \( p_{\text{Final}} \) into Equations (11) and (12) to obtain the simulated data \( d'_{\text{sim}} \). The discrepancy between the simulated \( d'_{\text{sim}} \) and observed seismic traces \( d_{\text{obs}} \) is quantified using the L2-norm, as shown in Equation (13), which guides the network to minimize the error during training and improves the accuracy of the reconstruction. The computed gradients are then used to simultaneously update all trainable parameters, including \( p_{\text{Adapt}} \) in the adaptive residual learning module and \( W \) in the U-Net network.

\begin{equation}
    w{'}_{}(x,z,t)=F(p{}_{Final})
\end{equation}

\begin{equation}
    d{'}_{sim}=S{}_{w{'}_{}}(x,z,t)
\end{equation}

\begin{equation}
\mathcal{O} = \frac{1}{2} \sum_{i=1}^{N} (d{'}_{sim} -d{}_{obs})^2
\end{equation}

\begin{equation}
p{}_{Adapt}^{j+1}=p{}_{Adapt}^{j}-\alpha \frac{\partial \mathcal{O}}{\partial p{}_{Adapt}^{j}}
\end{equation}

\begin{equation}
\label{e5}
W{}_{}^{j+1}=W{}_{}^{j}-\alpha \frac{\partial p{}_{}^{j}}{\partial W{}_{}^{j}  }\cdotp \frac{\partial \mathcal{O}}{\partial p{}_{}^{j}}
\end{equation}

\subsection{Implementation details}

In the algorithm implementation, the Pytorch framework is used in conjunction with the Deepwave package\cite{68}  to complete the computational workflow required for full-waveform inversion. The experiments are conducted on an NVIDIA GeForce RTX 4090 GPU. The optimization process employs the Adam algorithm, and the neural network weights are initialized using the Kaiming initialization method, with a learning rate of $5 \times 10{}^{-2}$. The weights of the adaptive residual network are initialized to zero, with a learning rate of $5 \times 10{}^{-1}$.

To validate the effectiveness of the proposed method, the Marmousi and Overthrust models were selected for inversion performance evaluation. These models exhibit complex structural features and significant velocity contrasts.

The Marmousi model is a widely used synthetic velocity model in geophysics. Known for its complex geological structure and velocity variations, it serves as a classic benchmark in seismic data processing and full-waveform inversion research. For the inversion validation, we downsampled the velocity model, resulting in a model size of (x × z) = (100 × 310) with a spatial grid increment of 0.03 km. The velocity range spans from 1,472 m/s to 4,000 m/s. The observed data and true model is shown in Fig.4 and 5 (b).

The Overthrust model is a geological model that simulates a geological environment with flat-lying sedimentary layers beneath the seabed. This model displays geological features of varying complexity, with a central region characterized by a typical thrust fault-fold belt, and the surrounding area consisting of monoclines and flat-lying sedimentary zones, effectively representing the diversity of complex geological conditions. The model has a size of (x × z) = (94 × 400), with a spatial increment of 0.03 km. The velocity range spans from 2,360 m/s to 6,000 m/s. The true model is shown in Fig 6 (b).

\begin{figure}[t]
	\centering
	\includegraphics[width=\linewidth]{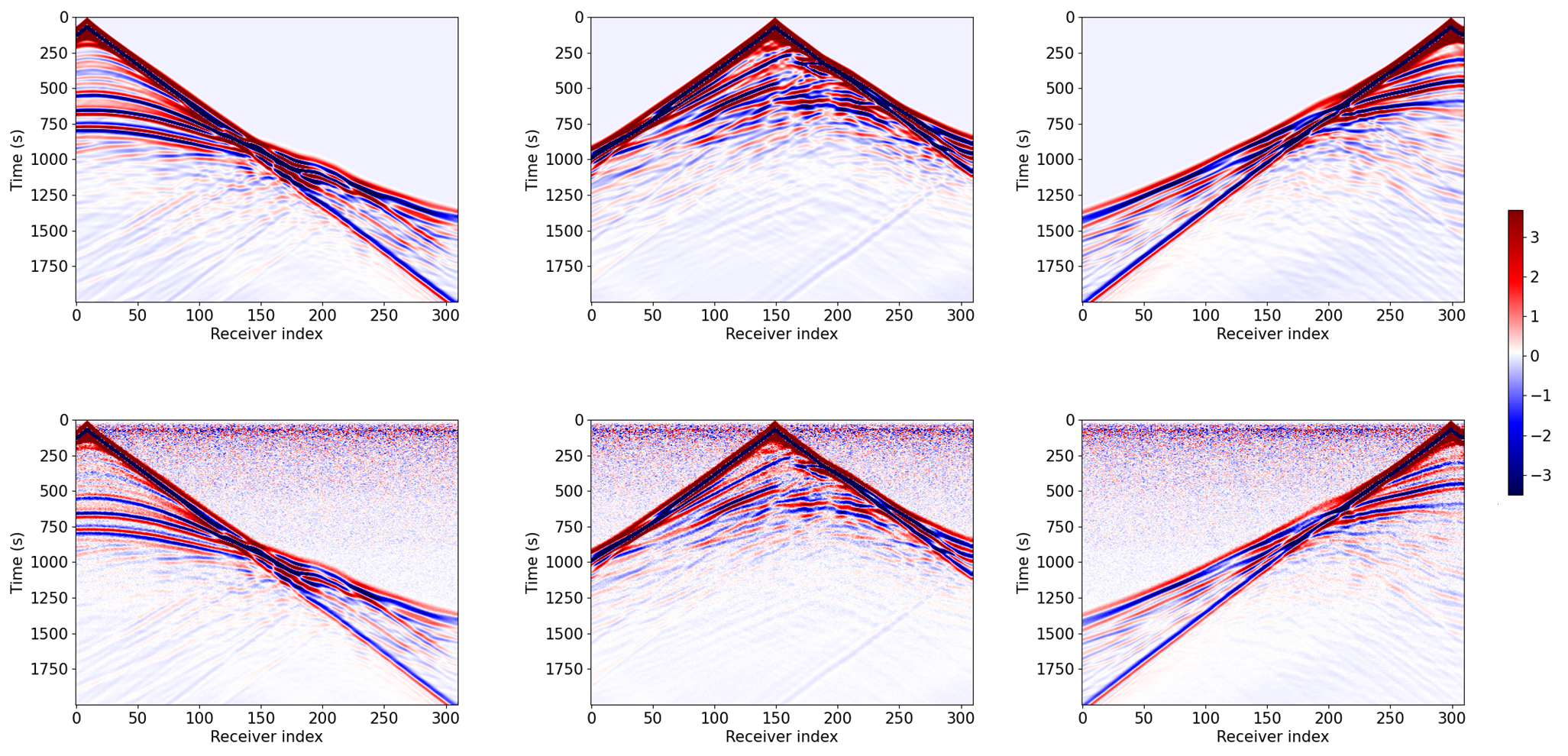}
	\caption{Three common-source shot gathers generated from the Marmousi model are shown. The first row represents forward data, while the second row corresponds to the same data with 10 dB additive Gaussian white noise. The introduction of noise serves to mimic the noisy conditions commonly encountered in real seismic data, enhancing the practical relevance of the study. Additionally, it provides a means to assess the inversion algorithm's robustness and its ability to withstand noise interference.}
\end{figure}

The receivers are configured with a fixed distribution geometry, with one receiver placed at each grid position. The Marmousi and Overthrust models are equipped with 310 and 400 receivers, respectively. Thirty uniformly spaced sources are positioned on the surface of both models, starting at the boundary of the model (0 meters at depth). The horizontal spacing for the two models is 0.3 km and 0.64 km, respectively. The source waveform used is a Ricker wave with a peak frequency of 5 Hz, and the data are recorded for 3 seconds with a sampling rate of 0.003 seconds. Therefore, in this acquisition configuration, the input to the encoder-decoder network consists of a 30-channel image with dimensions of 310×2000 or 400×2000, and the output is an image with dimensions of 100×310 or 94×400, corresponding to the size of the velocity model. The proposed method requires only 32,224,793 trainable parameters, demonstrating a lightweight design while maintaining high performance and offering a significant advantage in terms of model efficiency.

\begin{figure}[t]
	\centering
	\includegraphics[width=\linewidth]{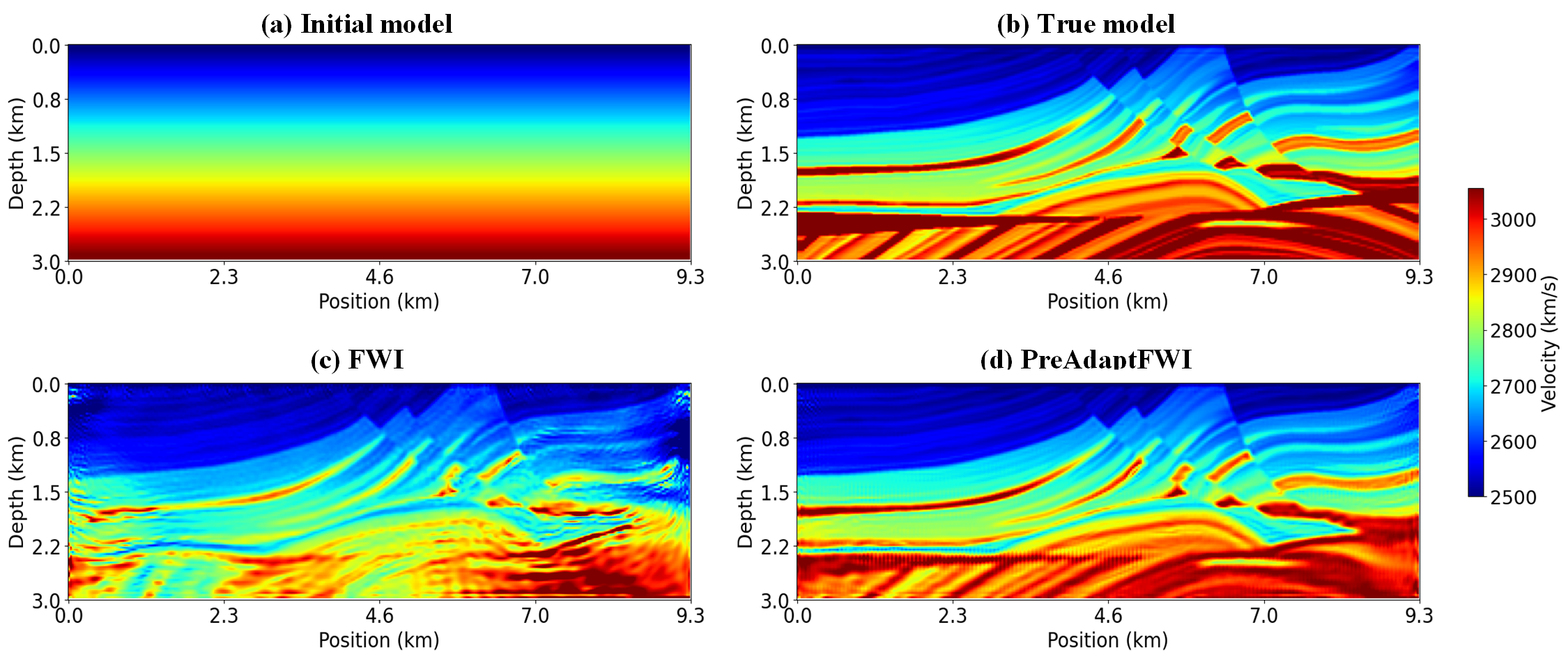}
	\caption{Comparison of the proposed method with traditional FWI on the Marmousi model: the first row shows the initial and true models, the second row compares inversion results under normal conditions.}
\end{figure}

\begin{figure}[t]
	\centering
	\includegraphics[width=\linewidth]{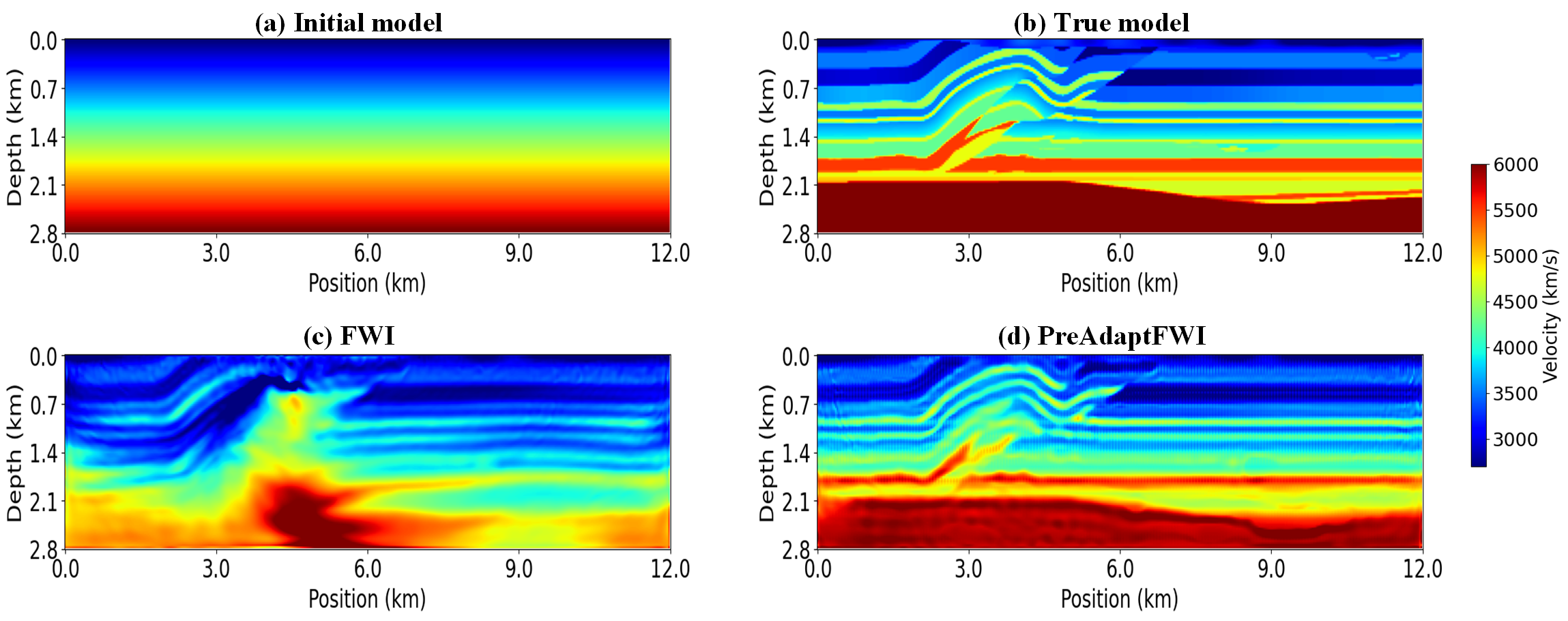}
	\caption{Comparison of the proposed method with traditional FWI on the Overthrust model: the first row shows the initial and true models, the second row compares inversion results under normal conditions.}
\end{figure}

\subsection{Pre-Training}
In the PreAdaptFWI framework, transfer learning is incorporated by training the Unet network using supervised learning. During the training process, the network parameters are adjusted by optimizing the objective function, which quantifies the error between the predicted velocity model and the true velocity model. The update of the network parameter gradients is based on the mean squared error (MSE) loss, which measures the difference between the predicted and true velocity models, without any feedback from the FWI gradients.

In recent years, the application of supervised learning in seismic inversion has largely drawn on methods from computer vision and natural language processing, leveraging the advantages of transfer learning by using a large and representative set of velocity models. However, due to the difficulty in acquiring real velocity models in seismic inversion, 

previous methods have relied on synthetic data\cite{52,65} or self-supervised techniques\cite{69,70,71} for training, which were effective exploratory approaches within the research context at the time and have yielded notable results. In this study, we propose a method that eliminates the need for manually generated synthetic training datasets, thereby saving time, reducing costs, and avoiding discrepancies between the features of synthetic and real data. By removing the dependency on synthetic data, the proposed method can better adapt to real geological conditions, enhancing the accuracy and reliability of seismic inversion results.

The main distinction of the supervised learning approach employed in PreAdaptFWI during pretraining, compared to previous studies, is that the neural network does not require manually collected or constructed seismic velocity models, which typically number in the dozens or even hundreds, for training. Instead, it is trained using a simple initial model provided by velocity analysis, ensuring stable network training. This enables the network to achieve acceptable inversion results as illustrated in Fig 5 (d) and Fig 6 (d). Therefore, the pretraining method proposed in this study overcomes the drawbacks of strong data dependency and training costs while retaining the advantages of efficiency and transferability. During the pretraining phase, the encoder-decoder network undergoes 3000 epochs of training using the Adam variant of the stochastic gradient descent algorithm, with a base learning rate set to $5 \times 10{}^{-2}$. In the subsequent training phase, the same input and output configurations as in the pretraining phase are used, eliminating the need for any modifications to the neural network. In this phase, the input data remains as 30 sources, with the output being the corresponding velocity model of the same size. 

The dataset-free pretraining method proposed in this study aims to stabilize the training of neural networks and accelerate the convergence of subsequent models to the correct solution. Furthermore, the method introduced in this paper demonstrates that even when using highly simplified uniform models, transfer learning can still significantly improve the training stability and convergence speed of the model. This, in turn, reduces the dependence on high-quality training data in practical applications and lays a foundation for the effective training of more complex models. In other words, 
the information required to construct the training set relies on minimal prior knowledge. This implies that we reduce our reliance on prior information, such as geological survey data of the studied region, including surface rock types, groundwater characteristics, the velocity distribution of different rock layers, and the overall velocity trend in the area, and instead, only need to make simple assumptions. Therefore, adopting this dataset-free pretraining method not only reduces the need for large amounts of labeled data and the introduction of prior knowledge, but also effectively enhances the performance of PreAdaptFWI in addressing real-world geological problems.

Although the Marmousi and Overthrust test models used in this study exhibit significant differences in geological features and velocity distributions, the velocity structure of the Marmousi model is more complex, with distinct sedimentary layers and irregular velocity variations, while the Overthrust model features a simpler layered structure with a relatively uniform velocity distribution. However, by applying the method proposed in this study and pretraining with a linearly distributed initial model, satisfactory inversion results were still achieved, as shown in Figure 5 and 6. This observation demonstrates that the proposed method does not require the creation of specific training sets for different models, thereby effectively enhancing the model's generalization capability. The method is adaptable to various geological backgrounds without relying on specific training data, significantly reducing data dependence and improving inversion efficiency.

\subsection{Adaptive residual learning}
Inspired by the concept of residual learning in computer vision \cite{72}, this study investigates an adaptive residual learning approach and applies it to the domain of full-waveform inversion. Figure 3 illustrates the differences between the proposed adaptive residual network and previous methods, providing a clear overview of the FWI framework combined with the adaptive residual module. Traditional convolutional modules (Section a in Figure 3), as fundamental components of neural networks, primarily extract features from input data through convolutional layers. In seismic data processing, the convolutional kernel slides over the data to perform convolution operations, capturing features at various scales and directions. Although effective in extracting fundamental waveform features from seismic wavefield data, these modules are limited when dealing with complex geological structures that have multi-scale and hierarchical features. Specifically, they struggle to capture both large-scale stratigraphic structures and finer details such as faults and folds simultaneously. Moreover, as the network depth increases, issues such as vanishing or exploding gradients may arise, affecting training efficiency and feature extraction capability. Traditional residual modules (Section b in Figure 3) are designed to address the vanishing gradient problem in deep convolutional networks. The core innovation of these modules lies in the introduction of identity mapping. In this module, the input data $X$ undergoes a series of convolution operations to generate $F(X)$, which is then added to the original input $X$ to produce the output. This structure helps to smooth the backpropagation of gradients during training, thereby accelerating the model's convergence. In the context of full-waveform inversion, residual modules help the network learn hierarchical features more effectively than traditional convolutional modules and improve the training of deep networks. However, their focus remains on learning global features, with limited capacity for capturing fine-grained local details, which restricts their effectiveness in delineating complex geological structures.

The adaptive residual module proposed in this study (Section c in Figure 3) represents a key innovation. This module consists of a weighted layer, a zeroing layer, and a parameter layer, enabling the neural network to not only learn the direct mapping from input $X$ to output $F(X)$, but also to specifically learn the residual mapping $F(X) - X$ through its unique structure. This innovation allows the proposed method to jointly optimize the background geological structure and velocity variations during the velocity model inversion process. Specifically, the U-Net framework employed hierarchically downsamples the input data through multiple convolutional modules and pooling layers. During this process, the receptive field of the convolution operations gradually increases, integrating information from larger regions and learning the direct mapping from input $X$ to output $F(X)$, thereby extracting more generalized features. These features capture large-scale velocity variations such as stratigraphic distribution, inter-layer velocity changes, and other global characteristics. Meanwhile, the adaptive residual learning module is applied in the final layer of the U-Net network. It takes the final output of the U-Net network as input and, through internal calculations, separates global features from detailed features, learning the residual mapping $F(X) - X$. This design allows for fine-tuning of the results after the network has processed large-scale features, focusing on capturing finer details such as geological structures, faults, and folds, which may have been overlooked during the extraction of large-scale features. Figures 17 and 18 illustrate the learning results, showing that by combining the two features learned by the U-Net and the adaptive residual learning module, the final result gradually approximates the true seismic velocity structure. Notably, unlike traditional residual modules, which are typically used after each convolution block, the adaptive residual learning module is applied only in the final layer of the network, thus avoiding unnecessary computational complexity and more effectively focusing on the fine-tuning of the inversion results. Furthermore, the proposed adaptive residual network is applicable to various CNN architectures, error evaluation functions, weight initialization schemes, and other variants, thereby enhancing optimization efficiency and the overall performance of the algorithm.

\begin{figure}[t]
	\centering
	\includegraphics[width=\linewidth]{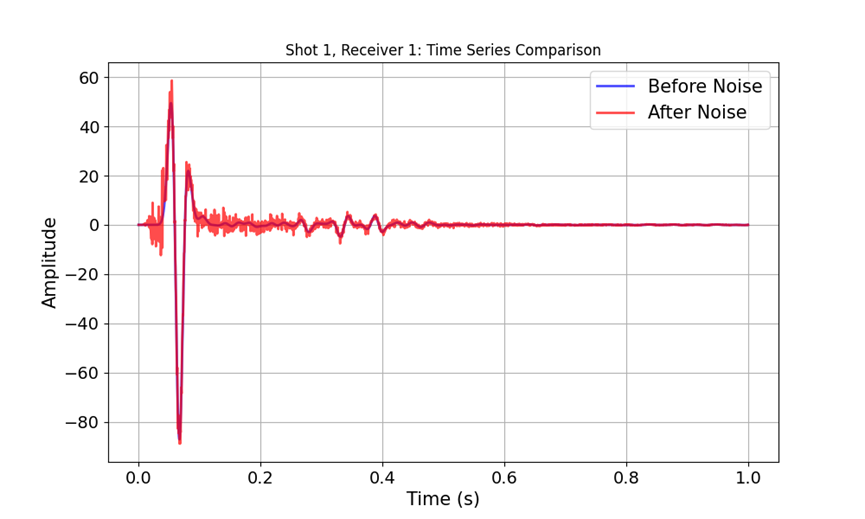}
	\caption{Comparison of the data before and after the addition of 10 db Gaussian white noise.}
\end{figure}

\section{Result}
To validate the effectiveness of the proposed method, we conducted extensive experiments on the Marmousi and Overthrust models. These experiments include: a comparison with traditional full-waveform inversion in terms of overall performance; evaluation under conditions of missing low-frequency information and added Gaussian white noise, compared with traditional methods; performance evaluation with different initial model, in comparison to traditional FWI results; and ablation studies on the pre-trained model and the adaptive residual module.

\begin{figure}[t]
	\centering
	\includegraphics[width=\linewidth]{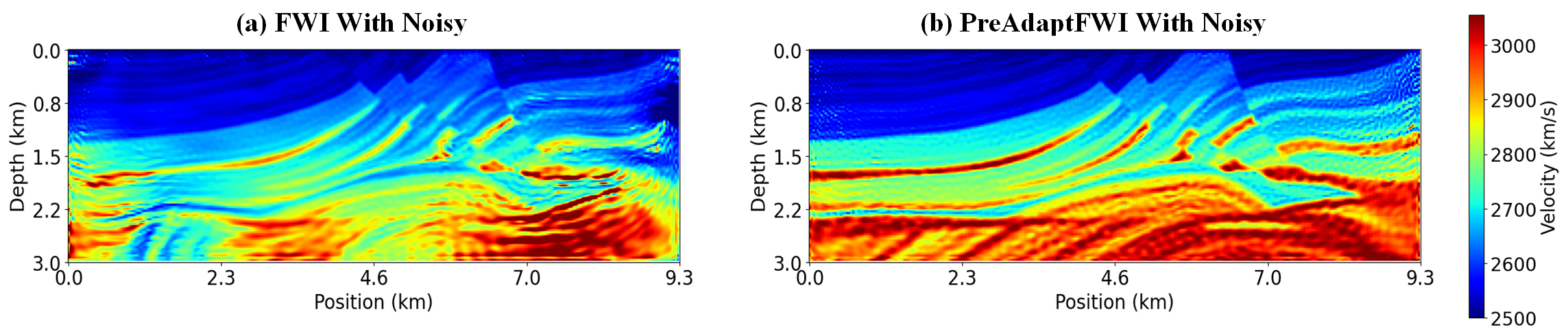}
	\caption{Comparison of the proposed method with traditional FWI on the Marmousi model: under the condition of 10 dB Gaussian white noise.}
\end{figure}

\begin{figure}[t]
	\centering
	\includegraphics[width=\linewidth]{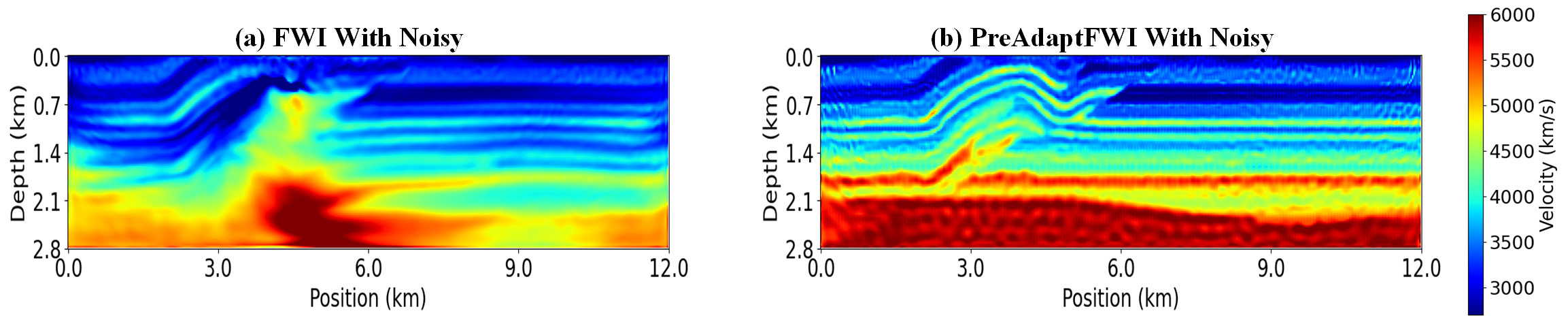}
	\caption{Comparison of the proposed method with traditional FWI on the Overthrust model: under the condition of 10 dB Gaussian white noise.}
\end{figure}

\begin{figure}[t]
	\centering
	\includegraphics[width=\linewidth]{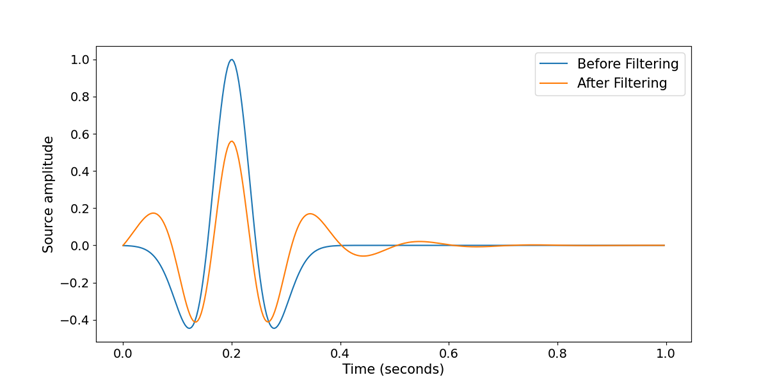}
	\caption{Comparison of source frequency spectra before and after filtering the 4 Hz low-frequency data.}
\end{figure}

\begin{figure}[t]
	\centering
	\includegraphics[width=\linewidth]{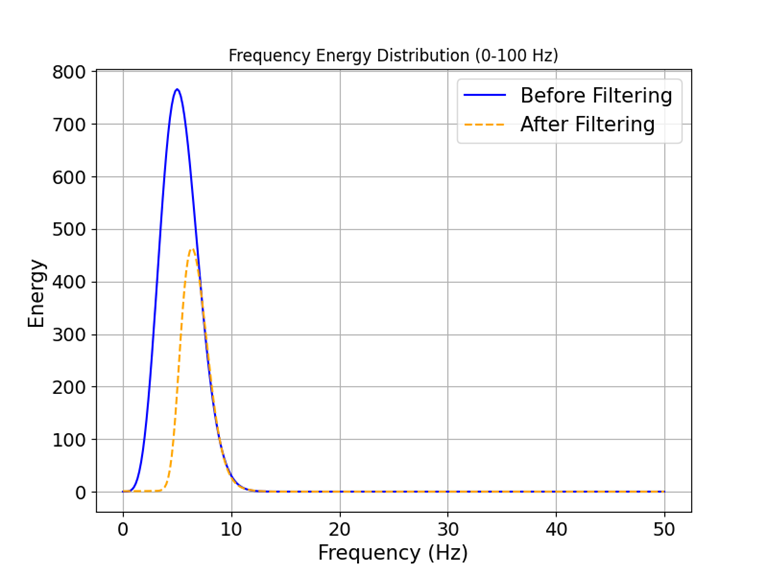}
	\caption{Comparison of the frequency energy distribution before and after filtering the 4 Hz low-frequency data.}
\end{figure}

\subsection{Testing under normal conditions}
The first row of Figure 4 presents three common-source gathers obtained from the Marmousi model, The second row of Figure 4 presents the corresponding source gathers after the addition of additive white noise, with the signal-to-noise ratio (SNR) of the noisy data set at 10 dB.

Figure 5 (a) illustrates a 1D linear initial model. In this case, due to the absence of the correct background velocity in the initial model, traditional full-waveform inversion  methods are prone to cycle skipping, as shown in Figure 5 (c), making it difficult to recover the true model (Figure 5 (b)). In contrast, Figure 5 (d) demonstrates the superior performance of the proposed PreAdaptFWI method in recovering the accurate geometric features of the Marmousi model. After 10,000 training iterations, each with 30 shot gathers, the neural network successfully captures geological features such as faults and anticlines. Additionally, we conducted experiments on the Overthrust model, achieving similarly satisfactory results, as shown in Figure 6 (d) compared to Figure 6 (b). These experimental results validate the superiority and universality of the PreAdaptFWI method. Specifically, PreAdaptFWI leverages a pre-training approach to optimize the model, enabling stable neural network outputs without relying on large amounts of labeled data, and more effectively adapting to real geological conditions. Furthermore, the introduced adaptive residual blocks allow the neural network to simultaneously learn the global stratigraphic distribution and inter-layer velocity variations, thereby capturing the detailed features of the geological model more comprehensively. These advantages make PreAdaptFWI a more robust and practical approach for full-waveform inversion tasks in complex geological environments, enhancing both the applicability and utility of the model.

\subsection{Testing with Gaussian white noise}
In practical applications, seismic data is often affected by various types of noise, which is typically unavoidable. Therefore, this study investigates the impact of Gaussian white noise with a signal-to-noise ratio of 10 dB, a common type of random noise, through corresponding experiments. The effect of noise on the trace is shown in Fig 7. By comparing the inversion results before and after the noise is added, we can assess the sensitivity of the proposed inversion method to noise and evaluate its ability to maintain accuracy in predicting the true subsurface structure. The experimental results, as shown in Fig 8 (b), indicate that, after 10,000 training iterations, the proposed PreAdaptFWI method is still able to recover the main geological features and fault positions, demonstrating its strong robustness and high predictive accuracy under noise interference. This suggests that PreAdaptFWI can effectively maintain accurate predictions of the true subsurface structure. The same noise level was also applied to the Overthrust model, with the results shown in Fig 9 (b). Again, PreAdaptFWI demonstrates strong predictive capability under noise influence. Compared to the traditional method in Fig 8 (a) and Fig 9 (a), PreAdaptFWI is able to better recover detailed geological features, proving the superiority and universality of this method in handling noise interference.

\subsection{Testing without the low-frequency information}
In seismic exploration, the absence of low-frequency data is a common phenomenon due to the limitations of measurement technology and the complexity of geological environments. In the absence of low-frequency data, the inversion process may suffer from cycle skipping problem. To assess the performance of the proposed method under such conditions, this study applies a 4 Hz low-frequency filtering cutoff to a Ricker wavelet 
with the dominant frequency of 5 Hz. The pre- and post-filtered data shown in Fig 10 and 11. Based on the processed data, the proposed method is able to gradually converge to a more accurate geological model, demonstrating better inversion stability and accuracy compared to the traditional method shown in Figure 12 (b). Furthermore, under the same low-frequency data loss conditions, inversion experiments were also conducted on the Overthrust model, as shown in Figure 13 (b). The results indicate that, compared to traditional methods, PreAdaptFWI is more effective in mitigating cycle skipping, maintaining the accuracy of inversion results, and accurately recovering key geological features, such as fault locations and inter-layer velocity variations, in the geological model. 

\begin{figure}[t]
	\centering
	\includegraphics[width=\linewidth]{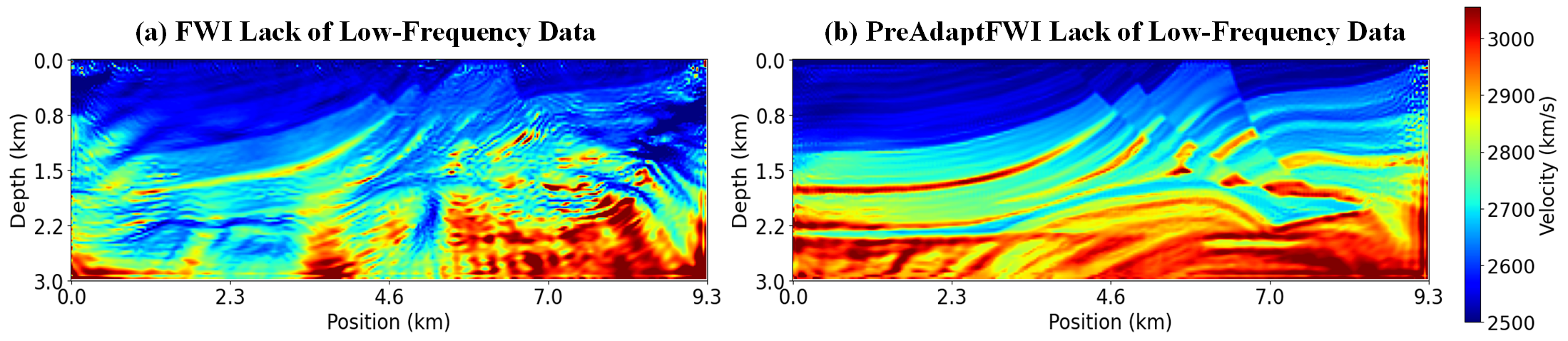}
	\caption{Comparison of the proposed method with traditional FWI on the Marmousi model: under the condition of a 4 Hz low-frequency filter cutoff and using a 5 Hz Ricker wavelet.}
\end{figure}

\begin{figure}[t]
	\centering
	\includegraphics[width=\linewidth]{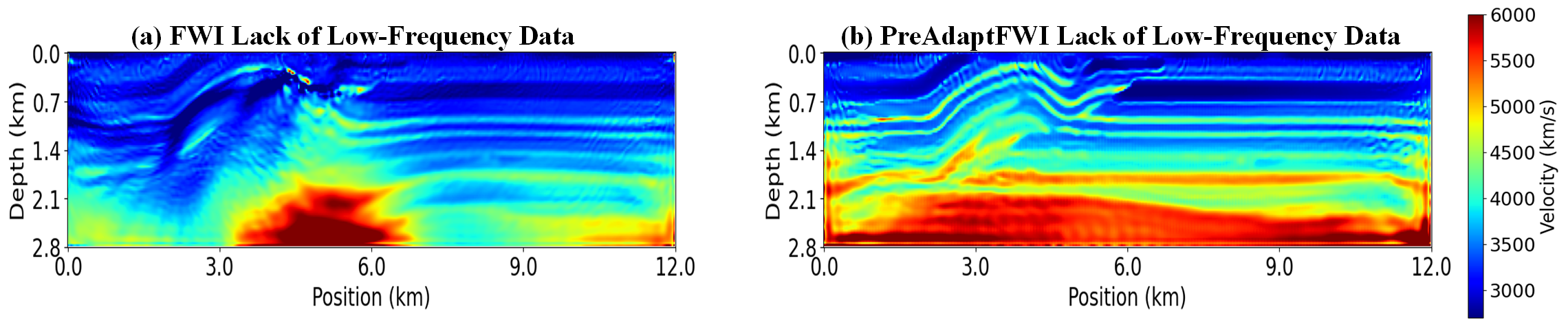}
	\caption{Comparison of the proposed method with traditional FWI on the Overthrust model: under the condition of a 4 Hz low-frequency filter cutoff and using a 5 Hz Ricker wavelet.}
\end{figure}

\begin{figure}[t]
	\centering
	\includegraphics[width=\linewidth]{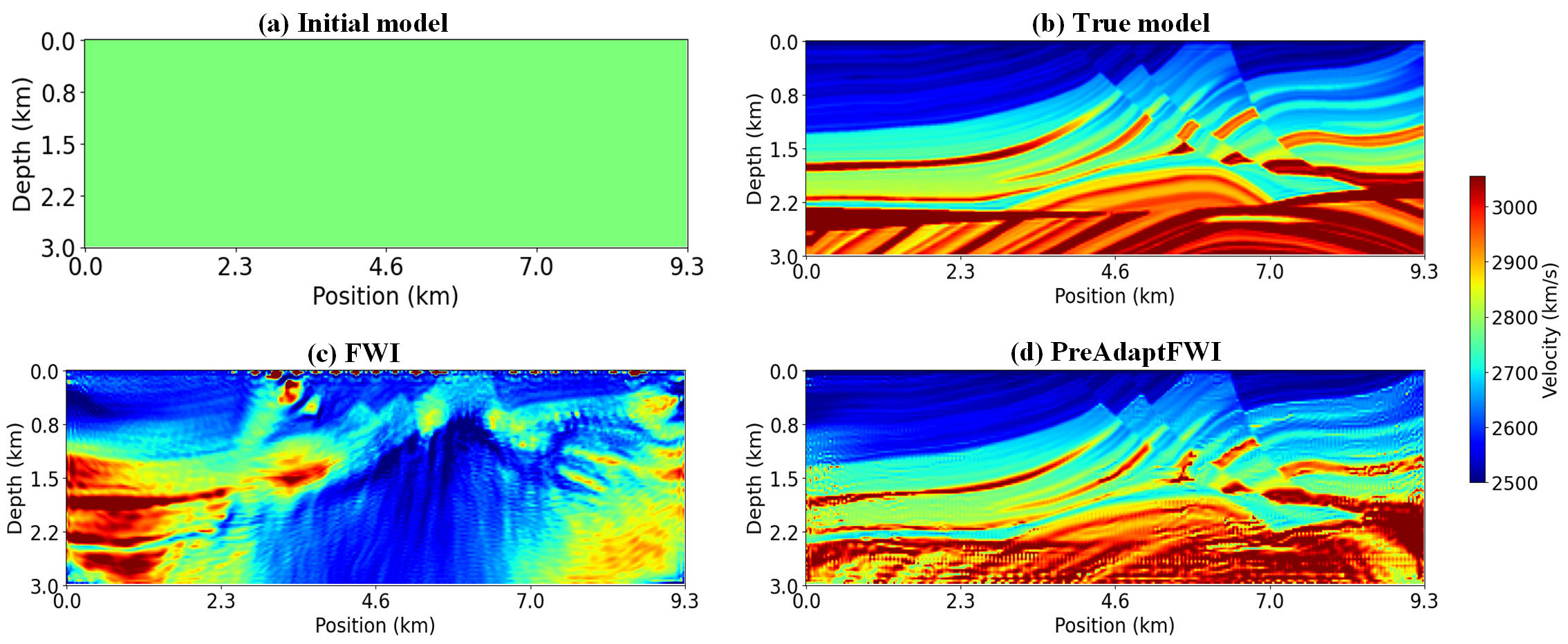}
	\caption{Comparison of the proposed method with traditional full-waveform inversion approaches on the Marmousi model using a uniform initial model: The first row shows the initial and true velocity models, while the second row presents the inversion results obtained by traditional FWI and the proposed method.}
\end{figure}

\begin{figure}[t]
	\centering
	\includegraphics[width=\linewidth]{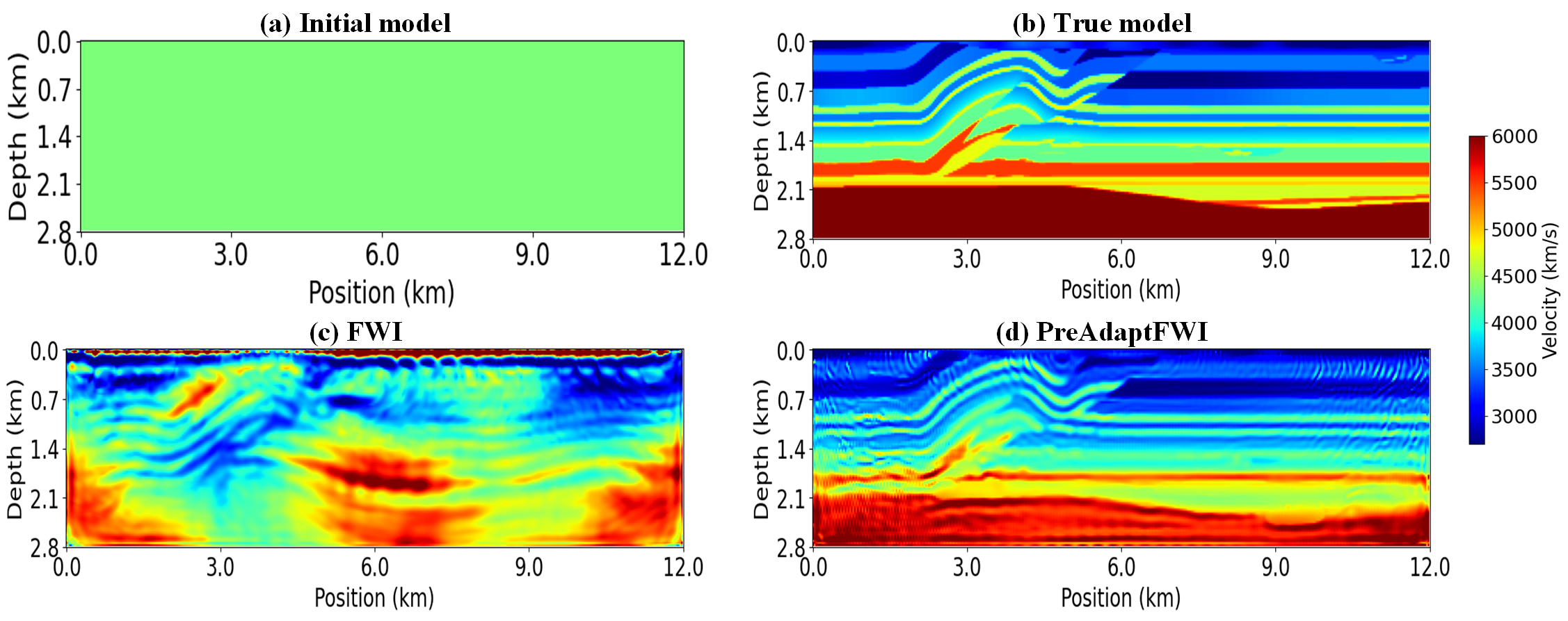}
	\caption{Comparison of the proposed method with traditional full-waveform inversion approaches on the Overthrust model using a uniform initial model: The first row shows the initial and true velocity models, while the second row presents the inversion results obtained by traditional FWI and the proposed method.}
\end{figure}

\subsection{Testing with the extremely poor initial model}
Full-waveform inversion can be viewed as an optimization problem, where the subsurface medium parameters are updated by minimizing the error between the simulated and observed waveforms. The initial model provides a starting point for the inversion process, helping the algorithm converge quickly to a relatively good solution. Without a suitable initial model, the inversion may fall into a local minimum or fail to converge. However, in practical measurements, the obtained seismic data are often incomplete and may be limited by factors such as topography, environment, and instrument precision. Data loss or distortion results in insufficient information for constructing the initial model. To evaluate the effectiveness of the proposed method in real-world applications, we conducted tests on the Marmousi and Overthrust models with a uniform initial model, as shown in Figures 14 (a) and 15 (a). The experimental results demonstrate that, compared to traditional full-waveform inversion methods that fail to converge to the correct solution, the proposed method successfully recovers a subsurface velocity structure close to the true solution under the uniform initial model conditions. This indicates that the method exhibits strong robustness and adaptability, effectively overcoming challenges posed by inaccurate initial models and enabling accurate model reconstruction in complex geological environments.
\subsection{Testing with the Mean Absolute Error quantitative evaluation metric}
To provide a comprehensive, objective, and standardized evaluation of the inversion performance of the proposed method, this chapter employs the Mean Absolute Error (MAE) as a quantitative assessment metric. MAE calculates the absolute differences between the model predictions and the true values, offering a clear representation of the inversion accuracy. Compared to other evaluation metrics such as MSE or RMSE, MAE is less sensitive to outliers, making it a more reliable indicator of the overall performance of the inversion results. The specific calculation formula is as follows:

\begin{equation}
MAE = \frac{1}{n}\sum\limits_{i = 1}^n {\left| {{P_i} - {T_i}} \right|}     
\end{equation}
Here, $P$ represents the inversion results inferred through the proposed method, and $T$ denotes the true velocity model. A lower MAE value indicates that the inversion results are closer to the true model, while a higher MAE value signifies the presence of larger errors. As shown in the comparison of experimental results from Tables 1 to 5, the MAE values of the proposed method are significantly lower than those of traditional methods, effectively demonstrating the significance of the PreAdaptFWI approach.

\begin{table}[H]
\centering
\caption{Comparison of Mean Absolute Error with Linear Initial Model in the Marmousi Model}
\begin{tabular}{ccc}
\hline
                      & FWI    & \multicolumn{1}{l}{PreAdaptFWI} \\ \hline
Linear initial model  & 274.01 & 125.21                          \\
With 10db Noisy       & 308.16 & 152.14                          \\
Lack of Low-Frequency & 485.82 & 196.33                          \\ \hline
\end{tabular}
\end{table}
\vspace{-23pt}
\begin{table}[H]
\centering
\caption{Comparison between Adaptive Residual Learning and Neural Network}
\begin{tabular}{ccccc}
\hline
\multicolumn{1}{l}{} & FWI    & NN Learning & Residual & PreAdapt \\ \hline
Marmousi             & 274.01 & 370.82      & 2812.07  & 125.21   \\
Overthrust           & 625.09 & 648.70      & 3824.32  & 143.73   \\ \hline
\end{tabular}
\end{table}
\vspace{-23pt}
\begin{table}[H]
\centering
\caption{Ablation Study Comparison on the Marmousi Model}
\begin{tabular}{cccc}
\hline
                          & Pre-training  & Residual Learning  & \multicolumn{1}{c}{Result} \\ \hline
\multirow{2}{*}{}         & \checkmark   & -            & 308.63                     \\ 
\multirow{1}{*}{Baseline} & -       & \checkmark        & 295.63                     \\ 
                          & \checkmark   & \checkmark        & 125.21                     \\ \hline
\end{tabular}
\end{table}
\vspace{-23pt}
\begin{table}[H]
\centering
\caption{Comparison of Mean Absolute Error with Extremely Poor Initial Model}
\begin{tabular}{ccc}
\hline
\multicolumn{1}{l}{} & FWI    & PreAdaptFWI \\ \hline
Marmousi             & 872.89 & 206.97      \\
Overthrust           & 935.81 & 182.08      \\ \hline
\end{tabular}
\end{table}
\vspace{-23pt}
\begin{table}[H]
\centering
\caption{Comparison of Mean Absolute Error with Linear Initial Model in the Overthrust Model}
\begin{tabular}{ccc}
\hline
\multicolumn{1}{l}{}  & FWI    & PreAdaptFWI \\ \hline
Linear initial model  & 625.09 & 143.73      \\
With 10db Noisy       & 688.23 & 147.66      \\
Lack of Low-Frequency & 586.43 & 213.75      \\ \hline
\end{tabular}
\end{table}

\begin{figure}[H]
	\centering
	\includegraphics[width=\linewidth]{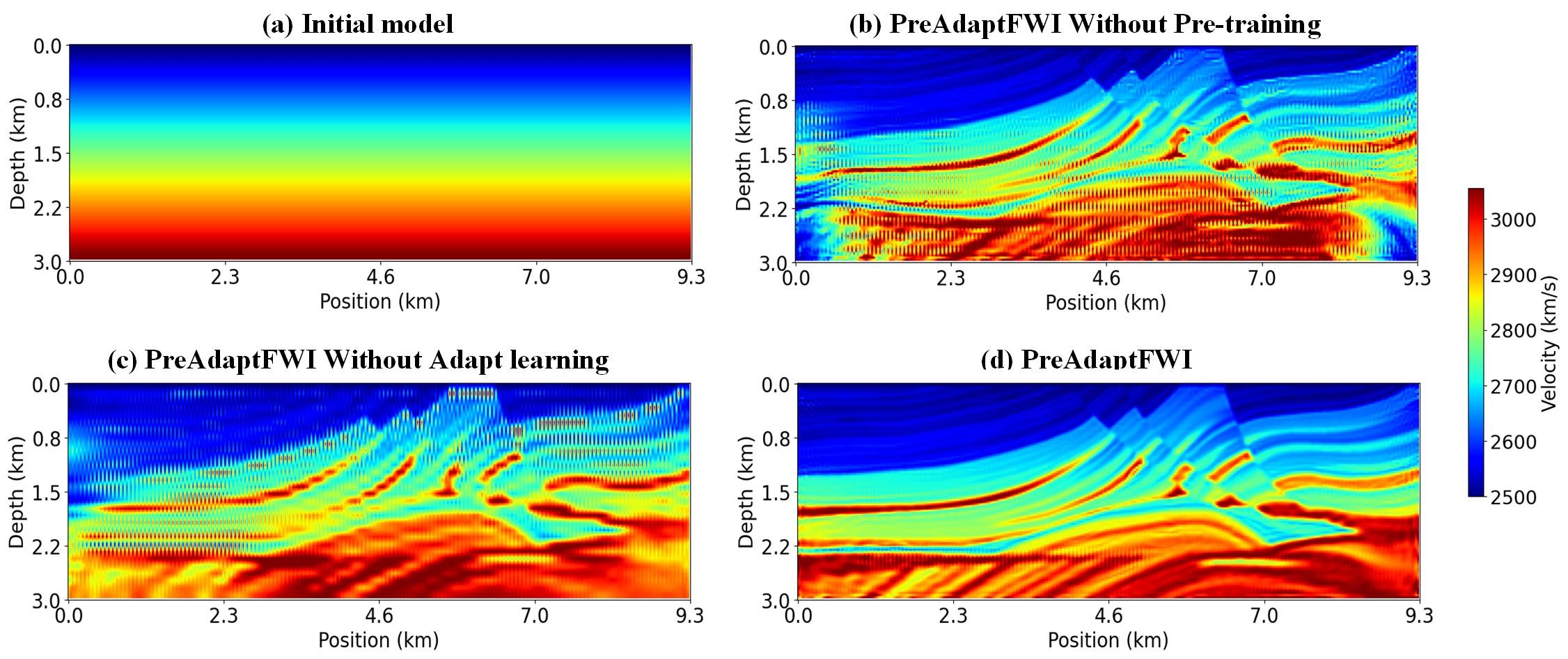}
	\caption{Ablation experiment results on the Marmousi model: The first row shows the effects of using a linear initial model and the adaptive residual learning module alone. The second row compares the results of using dataset-free pretraining alone with the combined use of the adaptive residual lea3rning module and dataset-free pretraining.}
\end{figure}

\begin{figure}[H]
	\centering
	\includegraphics[width=\linewidth]{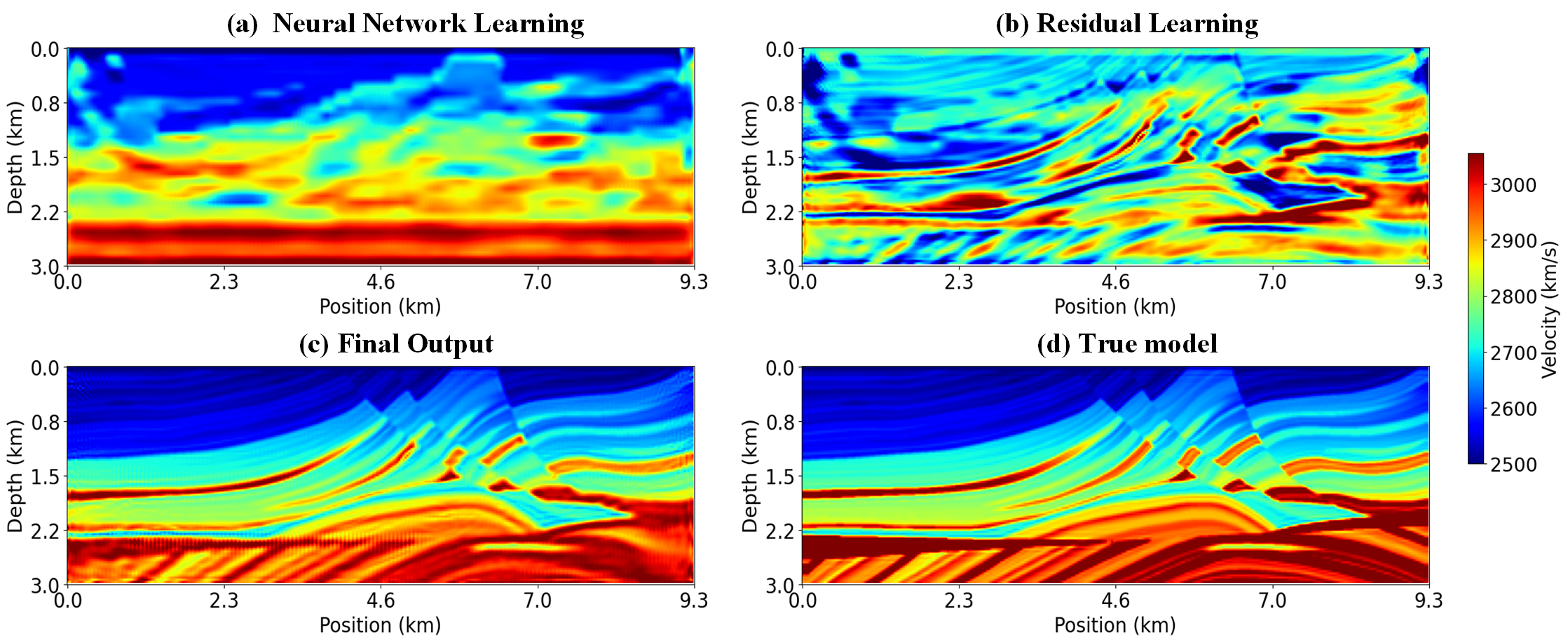}
	\caption{The Performance of the proposed adaptive residual learning module in conjunction with the neural network after 10000 epochs on the Marmousi model: The first row illustrates the features captured by the neural network and the adaptive residual learning module, respectively, while the second row compares the final output of the algorithm with the true velocity model.}
\end{figure}

\begin{figure}[H]
	\centering
	\includegraphics[width=\linewidth]{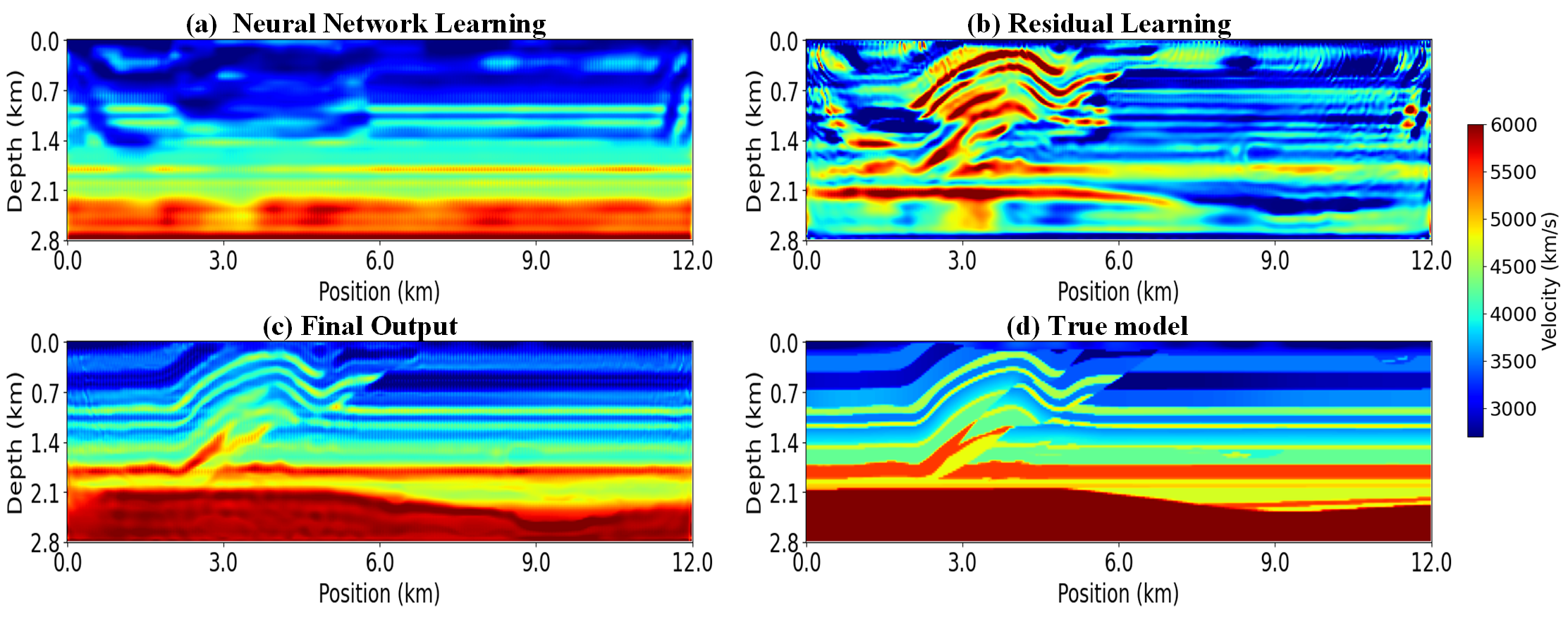}
	\caption{The Performance of the proposed adaptive residual learning module in conjunction with the neural network after 10000 epochs  on the Overthrust model: The first row illustrates the features captured by the neural network and the adaptive residual learning module, respectively, while the second row compares the final output of the algorithm with the true velocity model.}
\end{figure}

\subsection{Ablation study}
To validate the specific contribution of each innovation in the proposed method to the final model performance, we conducted tests on the Marmousi model, separately using pre-training with only the initial model and employing the model with only the adaptive residual learning block. The initial model used in the ablation experiments is shown in Fig 16 (a). As shown in Fig 16 (b), pre-training is crucial for stabilizing the neural network output, significantly reducing the appearance of artifacts and aiding in the recovery of deep velocity layers. The introduction of the adaptive residual learning block primarily reduces artifacts in the shallow velocity layers, as shown in Fig 16 (c). This indicates that the modules proposed in this study work synergistically during training, resulting in improved model performance. Furthermore, to explore what the neural network and adaptive residual learning block learn separately, we performed visualization experiments on both the Marmousi and Overthrust models, with results shown in Fig 17 (a) and (b), and Fig 18 (a) and (b). After introducing the adaptive residual learning block, the neural network focuses on the global distribution of geological layers and interlayer velocity variations, while the residual learning block focuses on capturing detailed features, including geological structures, faults, and folds. This collaborative strategy enhances overall performance, fully demonstrating the complementary nature of the two components.

\section{Conclusion}
This paper presents a framework that combines dataset-free pre-training with adaptive residual learning to assist in improving the performance of FWI, termed PreadAptFWI. Compared to prior approaches that rely on constructed synthetic datasets for training, the proposed method only requires moderate training on a simplistic initial model, thereby successfully overcoming the dependency on large amounts of labeled data. Furthermore, to enhance the model's performance, a universal adaptive residual learning module is introduced to collaboratively optimize the output of the neural network. Extensive experiments conducted on the Marmousi and Overthrust models demonstrate that PreAdaptFWI effectively mitigates the local minimum problem through the synergistic interaction between the neural network and the adaptive residual learning module, significantly improving the accuracy and stability of the inversion, regardless of whether low-frequency data, noise interference, or uniform initial model conditions are present. In future work, we intend to extend the PreadAptFWI framework to field seismic data, as the current validation has been limited to benchmark models. Adapting this method to field data presents additional challenges, including the need to address complex wave propagation effects, the uncertainty of source wavelets, and the presence of both coherent and random noise, which can significantly affect inversion accuracy. Overcoming these obstacles will be crucial for ensuring the robustness and general applicability of PreAdaptFWI in field seismic inversion scenarios.

\bibliography{main}
\bibliographystyle{IEEEtran}

\newpage

\end{document}